\input phyzzx.tex
\tolerance=1000
\voffset=-0.0cm
\hoffset=0.7cm
\sequentialequations
\def\rl{\rightline}

\def\t1{{\tilde 1}}

\def\t{\theta}

\REF{\SUSY}{Y. Shadmi and Y. Shirman, Rev. Mod. Phys. {\bf 72} (2000) 25, [arXiv:hep-th/9907225]; J. Terning, [arXiv:hep-th/0306119]; K. Intriligator and N. Seiberg, 
[arXiv:hep-th/0702069].}
\REF{\ISS}{K. Intriligator, N. Seiberg and D. Shih, JHEP {\bf 0604} (2006) 021, [arXiv:hep-th/0602239].}
\REF{\RAY}{S. Ray, Phys. Lett {\bf B642} (2006) 13, [arXiv:hep-th/0607172].}
\REF{\KIT}{R. Kitano, [arXiv:hep-ph/0606129]; Phys. Lett. {\bf B641} (2006) 203, [arXiv:hep-ph/0607090].}
\REF{\DM}{M. Dine and J. Mason, Phys. Rev. {\bf D77} (2008) 016005, [arXiv:hep-ph/0608063].}
\REF{\DIN}{M. Dine, J. L. Feng and E. Silverstein, Phys. Rev. {\bf D74} (2006) 095012, [arXiv:hep-th/0608159].}
\REF{\KOO}{R. Kitano, H. Ooguri and Y. Ookouchi, Phys. Rev. {\bf D75} (2007) 045022, [arXiv:hep-ph/0612139].}
\REF{\MUR}{H. Murayama and Y. Nomura, Phys. Rev. Lett. {\bf 98} (2007) 151803, [arXiv:hep-ph/0612186].}
\REF{\OS}{O. Aharony and N. Seiberg, JHEP {\bf 0702} (2007) 054, [arXiv:hep-ph/0612308].}
\REF{\CST}{C. Csaki, Y. Shirman and J. Terning, JHEP {\bf 0702} (2007) 099, [arXiv:hep-ph/0612241].}
\REF{\SHI}{K. Intriligator, N. Seiberg and D. Shih, JHEP {\bf 0707} (200) 017, [arXiv:hep-th/0703281].}
\REF{\HAB}{N. Haba and N. Mara, Phys. Rev. {\bf D76} (2007) 115019, arXiv:0709.2945[hep-th].}
\REF{\GIV}{A. Giveon and D. Kutasov, Nucl. Phys. {\bf B796} (2008) 25, arXiv:0710.0894[hep-th].}
\REF{\ZKKS}{A. Giveon, A. Katz, Z. Komargodski and D. Shih, JHEP {\bf 0810} (2008) arXiv:0808.2901[hep-th].}
\REF{\KOM}{Z. Komargodski and D. Shih, JHEP {\bf 0904} (2009) 093, [arXiv:hep-th/0902.0030].}
\REF{\KRD}{K. R. Dienes and B. Thomas, Phys. Rev. {\bf D78} (2008) 106011, arXiv:0806.3364[hep-th]; Phys. Rev. {\bf D79} (2009) 045001, arXiv:0811.3335[hep-th].}
\REF{\GK}{A. Giveon and D. Kutasov, Rev. Mod. Phys. {\bf 71} (1999) 983, [arXiv:hep-th/9802067].}
\REF{\SING}{F. Cachazo, K. Intriligator and C. Vafa, Nucl. Phys. {\bf B603} (2001), [arXiv:hep-th/0103067].}
\REF{\GEO}{C. Vafa, Journ. Math. Phys. {\bf 42} (2001) [arXiv:hep-th/0008142]; F. Cachazo, K. Intriligator and C. Vafa, Nucl. Phys. {\bf B603} (2001) 3, [arXiv:hep-th/0103067].}
\REF{\DINF}{E. Halyo, Phys. Lett. {\bf B387}  (1996) 43, [arXiv:hep-ph/9606423]; P. Binetruy and G. Dvali, Phys. Lett. {\bf B388}  (1996) 241, [arXiv:hep-ph/9606342].}
\REF{\EDI}{E. Halyo, JHEP {\bf 0407} (2004) 080, [arXiv:hep-th/0312042]; [arXiv:hep-th/0402155]; [arXiv:hep-th/0405269].}
\REF{\BGH}{I. Bena, E. Gorbatov, S. Hellerman, N. Seiberg and D. Shih, JHEP {\bf 0611} (2006) 088, [arXiv:hep-th/0608157].}
\REF{\FRA}{S. Franco, I. Garcia-Etxebarria and A.M. Uranga, JHEP {\bf 0701} (2007) 085, [arXiv:hep-th/0607218].}
\REF{\AMI}{A. Giveon and D. Kutasov, Nucl. Phys. {\bf B778} (2007) 129, [arXiv:hep-th/0703135]; JHEP {\bf 0802} (2008) 038, arXiv:0710.1833[hep-th].}
\REF{\ARG}{R. Argurio, M. Bertolini, S. Franco and S. Kachru, JHEP {\bf 0706} (2007) 017, [arXiv:hep-th/0703236].}
\REF{\BER}{M. Bertolini, S. Franco and S. Kachru, JHEP {\bf 0701} (2007) 017, [arXiv:hep-th/0610212].}
\REF{\OOG}{H. Ooguri and Y. Ookouchi, Phys. Lett. {\bf B641} (2006) 323, [arXiv:hep-th/0607183].}
\REF{\AGA}{M. Aganagic, C. Beem, J. Seo and C. Vafa, [arXiv:hep-th/0610249].}
\REF{\ABK}{M. Aganagic, C. Beem ad S. Kachru, Nucl. Phys. {\bf B796} (2008) 1, arXiv:0709.4277[hep-th].}
\REF{\OKS}{O. Aharony, S. Kachru and E. Silverstein, Phys. Rev. {\bf D76} (2007) 126009, arXiv:0708.0493[hep-th].}
\REF{\FU}{S. Franco and A. M. Uranga, JHEP {\bf 0606} (2006) 031, [arXiv:hep-th/0604136].}
\REF{\AHN}{C. Ahn, Class. Quant. Grav. {\bf 24} (2007) 1359, [arXiv:hep-th/0608160]; Class. Quant. Grav. {\bf 24} (2007) 3603, [arXiv:hep-th/0702038].}
\REF{\TAT}{R. Tatar and B. Wettenhall, Phys. Rev. {\bf D76} (2007) 126011, arXiv:0707.2712[hep-th].}
\REF{\VER}{M. Buican, D. Malyshev and H. Verlinde, JHEP {\bf 0806}(2008) 108, arXiv:0710.5519[hep-th].}

\singlespace
\rl{SU-ITP-09-25}
\pagenumber=0
\normalspace
\medskip
\bigskip
\titlestyle{\bf Metastable Vacua in Deformed N=2 Supersymmetric Models}
\smallskip
\author{ Edi Halyo{\footnote*{e--mail address: halyo@stanford.edu}}}
\smallskip
\centerline {Department of Physics} 
\centerline{Stanford University} 
\centerline {Stanford, CA 94305}
\smallskip
\vskip 2 cm
\titlestyle{\bf ABSTRACT}

We show that supersymmetric Abelian models that are obtained from deformations of those with ${\cal N}=2$ supersymmetry also contain metastable vacua for a wide range of
parameters. The deformations we consider are combinations of masses for charged and singlet fields, a singlet F--term and an anomalous D--term.
We find that, in all cases, the nonsupersymmetric vacua are parametrically far from the supersymmetric ones and therefore metastable.
Using retrofitting, we show that these metastable vacua may lead to semi--realistic phenomenology.

\singlespace
\vskip 0.5cm
\endpage
\normalspace

\centerline{\bf 1. Introduction}
\medskip

Realistic models of supersymmetry breaking have played a major role in the search for physics beyond the Standard Model for a long time. Until recently, supersymmetry
breaking was mainly investigated by building models with a unique nonsupersymmetric vacuum[\SUSY]. Models that preserve supersymetry were not explored since supersymmetric vacua
are the absolute minima of the scalar potential and it was assumed that the models would always relax to these vacua. However, in principle, the existence of a supersymmetric vacuum does
not necessarily exclude the possibility of a local minimum of the scalar potential which breaks supersymmetry. Indeed, recently it was realized that there is a large class of models that have 
nonsupersymmetric local minima in addition to supersymmetric vacua. If these local minima are stable in all field directions and very far from the supersymmetric ones in field space,
then they are metastable and may lead to semi--realistic phenomenology. 


After ref. [\ISS], many models with metastable nonsupersymmetric vacua have been constructed[\RAY-\KOM]. These models mostly have non--Abelian gauge groups and therefore strong interactions 
that break supersymmetry dynamically. In this letter, we consider a simpler possibility, metastable nonsupersymmetric vacua in models with an Abelian gauge group and simple 
perturbative superpotentials.{\footnote1{For previous work in which metastable vacua arise perturbatively see [\KRD].}} 
More concretely, we consider Abelian models with ${\cal N}=1$ supersymmetry which are obtained from those with ${\cal N}=2$ supersymmetry by all possible
deformations. These deformations include mass terms for the chiral multiplets, F--terms, anomalous D--terms and mass terms for the singlets which break supersymmetry from
${\cal N}=2$ to ${\cal N}=1$. We find that, in these models, metastable supersymmetry breaking vacua are quite generic.  

The basic model we consider has ${\cal N}=1$ supersymmetry and an Abelian gauge group. The matter sector contains the hypermultiplet of the ${\cal N}=2$ supersymmetric theory which in the
${\cal N}=1$ notation consists of two oppositely charged chiral fields. In addition, there is a gauge singlet which is the scalar in the ${\cal N}=2$ gauge multiplet. These couple
through a superpotential that includes a Yukawa term inherited from the ${\cal N}=2$ supersymmetric theory.
Two of the deformations are mass terms for the charged chiral fields and the singlet. In addition, we can include two more deformations: an F--term for the singlet
and/or an amomalous D--term. We show that these simple models with no strong dynamics have nonsupersymmetric vacua in addition to supersymmetric ones for a wide range of the parameters. 
Supersymmetry is broken at tree level by either F or D-terms. In both cases the supersymmetric and nonsupersymmetric vacua are parametrically far from each other; therefore the 
nonsupersymmetric vacua can be made metastable by a suitable choice of parameters. 

Since supersymmetry is broken at tree level and not dynamically, the supersymmetry breaking scale is not suppressed. However, this can be accomplished by retrofitting; i.e. by coupling our
models to hidden non--Abelian gauge groups by nonrenormalizable terms. In this case,
we estimate the soft supersymmetry breaking masses in the metastable vacua and show that, using retrofitting, they may lead to semi--realistic phenomenology.

In addition to their simplicity, another advantage of these models is the fact that they can be easily embedded in larger ones such as models with a $U(N_c)$ gauge group 
with $N_f$ pairs of fundamentals and an adjoint. Specializing to the Abelian subgroup of $U(N_c)$ and setting the VEVs of all the fundamentals except for one pair to zero, the 
non--Abelian model reduces to ours. Our results would be especially relevant at high energies where the physics is weakly coupled and one can neglect the non--Abelian dynamics.  

Since these models are deformed ${\cal N}=2$ supersymmetric models, they can be easily realized on D--branes, for example either on intersecting
branes[\GK] or on branes wrapped on singularities[\SING]. In both cases, the construction and deformations of the ${\cal N}=2$ supersymmetric theory are well--known. In fact, using the
D--brane picture, our model can be easily generalized in many directions resulting in a large class of models with metastable nonsupersymmetric vacua.

This paper is organized as follows. In Section 2 we describe the models with nonzero F or anomalous D--terms and find their supersymmetric and metastable vacua. 
In Section 3 we show that, using retrofitting, the metastable vacua of the models may
lead to semi--realistic phenomenology. Section 4 contains our conclusions and a discussion of our results.

\bigskip
\centerline{\bf 2. Models with Metastable Supersymmetry Breaking Vacua}
\medskip

In this section, we describe some simple supersymmetric models of the type described above with metastable nonsupersymmetric vacua. We consider two types of models: 
those that break supersymmetry by nonzero F or D--terms. 
In both cases, supersymmetry breaking is at tree level. We find that even for these relatively simple models metastable vacua are
quite generic, i.e. they exist for a wide range of the parameters.

{\it 2.1. Models with F--term Supersymmetry Breaking}:
Consider a model with ${\cal N}=1$ supersymmetry and a $U(1)$ gauge group. The matter sector consists of two charged fields $q_1,q_2$, (with charges $\pm 1$) and a neutral  
field $\Phi$ with the superpotential 
$$W=\lambda \Phi q_1 q_2 + m q_1 q_2+ M \Phi^2+ F \Phi \eqno(1)$$ 
We included mass terms for both $\Phi$ and $q_1,q_2$ to make the superpotential as general as possible. In fact, the only term that we neglected but is allowed by the local $U(1)$ 
symmetry is a $\Phi^3$ term which would not change our results qualitatively. As mentioned in the introduction, this model can be obtained from 
a generic deformation of a model with ${\cal N}=2$ supersymmetry (which would include only the Yukawa term). The deformations are the two mass terms and the F--term. Note that
here we set the remaining deformation, i.e. the anomalous D--term to zero.
The scalar potential is given by
$$V_F=|F_{\Phi}|^2+|F_{q_1}|^2+|F_{q_2}|^2 \eqno(2)$$
where
$$F_{\Phi}=\lambda q_1 q_2+ 2M \Phi+ F \eqno(3)$$
$$F_{q_1}=(\lambda \Phi+m)q_2 \eqno(4)$$
and
$$F_{q_2}= (\lambda \Phi +m)q_1 \eqno(5)$$
In addition, there is the $U(1)$ D--term contribution to the scalar potential
$$V_D=g^2(|q_1|^2-|q_2|^2)^2 \eqno(6)$$
The total scalar potential is the sum $V=V_F+V_D$.
Since our aim is only to show the existence of metastable vacua in these models, our analysis of the vacua will not be complete. In particular, as a simplification, we will 
consider only the D--flat directions with either $q_1=q_2$ or $q_1=-q_2$. Note that D--flatness requires only that $|q_1|=|q_2|$ where the phases of $q_1,q_2$ are not fixed.

{\it 2.1.1. The case with $q_1=q_2$}:
We take $q_1=q_2$ in order to satisfy the D--term condition. Then we find two supersymmetric vacua at
$$q_1=q_2=0 \qquad \Phi=-{F \over {2M}} \eqno(7)$$ 
and
$$q_1=q_2=\pm \sqrt{{{2mM} \over \lambda^2}- {F \over \lambda}} \qquad \Phi=-{m \over \lambda} \eqno(8)$$

Under the above assumptions, there are nonsupersymmetric minima of the scalar potential at (For details see the Appendix.)
$$q_1^2=q_2^2={M \over {\lambda^2}}(\lambda \Phi+ m) \eqno(9)$$
where
$$\Phi={1 \over {2 \lambda^2}}[-(3 \lambda M+2 \lambda m) \pm \sqrt{(3 \lambda M+ 2 \lambda m)^2-4 \lambda^2(m^2+\lambda F+mM)}] \eqno(10)$$

Due to the complexity of the above VEVs it is hard to check the metastability of the vacua in eqs. (9) and (10). Therefore, in order to simplify the analysis, we consider these vacua
in certain limits of the parameters which correspond to particular corners of the parameter space.

In the limit $M>>m>>\sqrt {F}$, eqs. (9) and (10) describe two pairs of vacua given by 
$$\Phi=-{{3M} \over \lambda} \qquad q_1=q_2=\pm i {\sqrt 3 \over \lambda}M \eqno(11)$$ 
and 
$$\Phi=-{m \over {3 \lambda}} \qquad q_1=q_2=\pm {\sqrt{2 \over 3}} {\sqrt{mM} \over \lambda} \eqno(12)$$
Both pairs of vacua break supersymmetry since all F--terms given by eqs. (3)--(5) are nonzero there. In addition, we need to check the local stability of the vacua in eqs. (11) and (12) since
in the limit $M>>m>>\sqrt{F}$ they may become unstable. This is done by checking that all scalar masses squared are positive at these vacua (Details are given in the Appendix).
From the form of the scalar potential it is easy to see that all three scalar fields $\Phi,q_1,q_2$ mix with each other. An analysis 
of the mass squared matrix shows that the vacuum in eq. (11) (eq. (12)) is not locally stable in this limit since this requires $|q|^2<0$ ($\lambda F>2mM$ which contradicts the above limit). 


In another limit, e.g. when $m>>M>>\sqrt{F}$ the vacua in eq. (9) and (10) reduce to
$$\Phi=-{{m} \over \lambda} \pm {\sqrt{2} \over \lambda} \sqrt{mM}  \qquad q_1^2=q_2^2={\sqrt{2} \over \lambda^2} \sqrt{mM^3}  \eqno(13)$$
It is easy to show that these vacua are locally stable if we choose $M>0$. We note that these vacua do not depend on $F$ so that we could have set $F=0$ in the superpotential. 
It is safe to neglect one--loop effects since these vacua are fixed at tree level. If $M=0$, the only vacuum is the supersymmetric one given by $q_1=q_2=0$ and a free $\Phi$. In this case, 
there are no loop corrections since supersymmetry in not broken and $\Phi$ is a real modulus. 

We note that all VEVs in the nonsupersymmetric vacua are inversely proportional to $\lambda$, and therefore parametrically far from the supersymmetric ones. As a result,
for small enough $\lambda$, these nonsupersymmetric vacua are metastable. Clearly, when $\lambda \to 0$ they (together with the supersymmetric 
one in eq. (8)) escape to infinity and disappear. 

In a third limit of the parameters, i.e. when $\sqrt{F}>>M>>m$, we find the nonsupersymmetric vacua
$$\Phi=\pm \sqrt{-{F \over \lambda}} \qquad q_1=q_2=\pm \left(-{{FM^2} \over \lambda^3} \right)^{1/4} \eqno(14)$$
where the VEVs are real only if $F<0$. It can be shown that these vacua are locally stable for $\lambda |F|<16M^2$ which requires a small $\lambda$. We see that they are also
parametrically far from the supersymmetric ones, since the VEVs above are proportional to inverse powers of $\lambda$. Thus, these vacua can also be made metastable for small enough $\lambda$.
Note that in this case neither vacuum depends on $m$; therefore we conclude that these metastable vacua exist even for $m=0$. If $M=0$, at tree level, we get a nonsupersymmetric vacuum 
with $q_1=q_2=0$ and a flat direction, parametrized by the pseudomodulus $\Phi$. The one--loop potential for $\Phi$ has a minimum at $\Phi=0$. However, this vacuum is not locally stable
since there is a tachyonic direction at the origin of the field space. In this case, the only vacuum is the supersymmetric one at $\Phi=0$ and $q_1=q_2=\sqrt{-F/\lambda}$.


Usually, in models with metastable supersymmetry breaking vacua, it is the supersymmetric vacuum that is parametrically far from the origin of field space where one finds 
the nonsupersymmetric vacuum. However,
this is guaranteed only in models with generic superpotentials (which contain all the terms that are allowed by the global symmetries) and a small R violating term. Above we found just 
the opposite; the nonsupersymmetric vacua are parametrically far from the origin. This should not be surprising because the superpotential in eq. (1) is not generic for any assignment of 
R charges. There is no R charge assignment for $\Phi$ for which both the mass and F--terms are invariant and the Yukawa term is a small R breaking correction.

{\it 2.1.2. The case with $q_1=-q_2$}:
We can satisfy the D--term constraints also by taking $q_1=-q_2$. The analysis is very similar to the case above and gives the supersymmetric vacua
$$\Phi=-{m \over \lambda} \qquad q_1=-q_2=\pm \sqrt{{F \over \lambda}-{{2Mm} \over \lambda^2}} \eqno(15)$$
which is similar to the supersymmetric solution in eq. (8). In this case, clearly there is no solution with $q_1=q_2=0$. 
We see that these vacua are parametrically far from the origin of field space for small $\lambda$. The nonsupersymmetric metastable vacua are at (with $q_1=-q_2=q$)
$$q^2=-{M \over \lambda^2}(\lambda \Phi+m) \eqno(16)$$
where $\Phi$ is fixed by
$$\Phi={1 \over {2 \lambda^2}}[-(2 \lambda m-3 \lambda M) \pm \sqrt{(2 \lambda m- 3 \lambda M)^2-4 \lambda^2(m^2-\lambda F-mM)}] \eqno(17)$$
which differs from eq. (9) only by a few signs. As above, we can take different limits of parameters to explore these vacua in particular corners of the parameter space.
In the limit $M>>m>> \sqrt{F}$, there are two pairs of vacua given by
$$\Phi={{3M} \over \lambda} \qquad q=\pm i {\sqrt{3} \over \lambda}M \eqno(18)$$
and
$$\Phi=-{m \over {3 \lambda}} \qquad q\pm i \sqrt{2 \over 3} \sqrt{mM} \eqno(19)$$
The analysis of the stability of the above pairs of vacua is precisely the same as those in eqs. (11) and (12). Therefore, the vacua in eqs. (18) and (19) are not locally stable. 

In the limit $m>>|M|>>\sqrt{F}$, we find the nonsupersymmetric vacua with ($q_1=-q_2=q$)
$$\Phi=-{{m} \over \lambda} \pm {\sqrt{2} \over \lambda} \sqrt{m|M|}   \qquad q^2= {\sqrt{2} \over \lambda^2} \sqrt{m|M|^3}  \eqno(20)$$
An analysis similar to the one for the vacua in eq. (13) shows that these vacua are locally stable for $M<0$. Finally, if $\sqrt{F}>>M>>m$, we find the vacua
$$\Phi=\pm \sqrt{{F \over \lambda}} \qquad q_1=-q_2=\pm  \left({{FM^2} \over \lambda^3} \right)^{1/4} \eqno(21)$$
which are locally stable for $\lambda F<16M^2$.

We see that for $q_1=-q_2$ both the supersymmetric and nonsupersymmetric vacua are given by VEVs that are proportional to inverse powers of $\lambda$, so all vacua are parametrically far
from the origin of field space. However, they are also parametrically far from each other and therefore eqs. (20) and (21) describe metastable nonsupersymmetric vacua. 
The discussion of one--loop effects on the above vacua are similar to those in the previous section and will not be repeated here.

{\it 2.2. Models with D--term Supersymmetry Breaking}:
We now describe the case in which supersymmetry is broken by a nonzero D--term in the metastable vacuum. For D--term breaking we consider the above model with the superpotential
$$W=\lambda \Phi q_1 q_2 + m q_1 q_2+ M \Phi^2  \eqno(22)$$ 
i.e. eq. (1) with $F=0$. The F--terms and $V_F$ are given by eqs. (3)-(6) as before.
On the other hand, we now assume that there is an anomalous D--term $\xi$ so that
$$V_D=g^2(|q_1|^2-|q_2|^2+ \xi)^2 \eqno(23)$$
The total scalar potential is again $V=V_F+V_D$.

In this case, the metastable nonsupersymmetric vacuum is located at the origin of the field space, i.e. at $\Phi=q_1=q_2=0$ where 
supersymmetry is broken since $D=\xi \not =0$. This vacuum is locally stable if all the scalar masses squared are positive at the origin which is
guaranteed if $m_{q_2}^2=m^2-2g^2 \xi>0$ (since the masses squared for the other scalars are always positive). 

The supersymmetric vacua which are found by setting the $F$ and $D$ terms to zero are given by
$$\Phi=-{m \over \lambda} \qquad q_1={{2Mm} \over {\lambda^2 q_2}} \eqno(24)$$
where
$$q_2^2={\xi \over 2} \pm {1 \over {2 \lambda^2}} \sqrt{\lambda^4 \xi^2+ 16 M^2m^2} \eqno(25)$$
We see that all the VEVs in eqs. (24) and (25) are 
proportional to inverse powers of $\lambda$. Therefore, these supersymmetric vacua are parametrically far from the origin and the nonsupersymmetric vacuum at the origin can be made 
metastable by choosing a small enough $\lambda$.

The location of the supersymmetric vacua far from the origin can be explained in terms of R symmetry as usual. We can assign the R charges $R[q_1]=R[q_2]=R[\Phi]=1$ under which 
the mass terms in the superpotential in eq. (1) are invariant whereas the Yukawa term is not. If $\lambda<<1$, then the Yukawa term is a small perturbation
that breaks R symmetry and restores supersymmetry far from the origin (since the supersymmetric vacuum comes from infinity). This is different from what we found in section 2.1.1
since in the presence of a nonzero F--term, there is no assignment of R charges under which the superpotential (for $\lambda=0$) is R invariant. 

In the case of D--term supersymmetry breaking, it is crucial that the two masses $m$ and $M$ and the anomalous D--term $\xi$ are all nonzero. If $m=0$, the origin is not stable and therefore 
not a vacuum. On the other hand, if $M=0$, then $\Phi$ is classically a flat direction. It gets a potential at the one--loop level which has a minimum at the origin, $\Phi=0$. It can be 
shown that this is a metastable vacuum for small enough $\lambda$. Finally, if $\xi=0$ the origin is a minimum but also supersymmetric.
This should be contrasted with the F--term breaking cases above in which we could set some of the parameters to zero and still have a metastable vacuum. 

As we mentioned above, these models may have other supersymmetric and nonsupersymmetric vacua since our analysis was not complete due to our assumption that D--flatness is satisfied
by either $q_1=q_2$ or $q_1=-q_2$. 
Our aim in this letter is simply to show the possibility of obtaining metastable nonsupersymmetric vacua in these models and our results should be taken as an existence proof.

\bigskip
\centerline{\bf 3. Retrofitting and Phenomenology}
\medskip

Since the models we considered above break supersymmetry at tree level they cannot describe dynamical supersymmetry breaking at an exponentially small energy scale. This can be 
accomplished by retrofitting i.e. by adding 
nonrenormalizable terms to the superpotential which describe the embedding of our models into larger ones at high energies. For example, the dimensionful parameters $m,M$ and $F$ can be 
obtained from the nonrenormalizable terms
$$\int d^2 \theta~ Tr(W_{\alpha})^2 \left(c_1 {{q_1 q_2} \over M_P^2} + c_2 {\Phi \over M_P} + c_3 {\Phi^2 \over M_P^2} \right) \eqno(26)$$
where $c_i \sim O(1)$ are model dependent coefficients and
$W_{\alpha}$ describes the gauge superfield of a new non--Abelian gauge group which becomes strongly coupled at a high energy scale $\Lambda$. Gaugino condensation of this group gives
$<Tr(W_{\alpha})^2> \sim \Lambda^3$ and as a result we obtain the masses $m \sim c_1 \Lambda^3/M_P^2$, $M \sim c_3 \Lambda^3/M_P^2$ and the F--term $F \sim c_2 \Lambda^3/M_P$. We see that
$F>>m \sim M$. This is a result of the fact that we used the same non--Abelian group in eq. (26) for all the nonrenormalizable terms. However, if we want to obtain, for example,
$m>>M>> \sqrt{F}$ as we assumed above for simplicity, we have to introduce three non--Abelian gauge groups, one for each term with different scales $\Lambda_{1,2,3}$. The case with $F=0$ can be 
obtained by assuming that $\Phi$ is odd under a discrete symmetry which prevents the second term in eq. (26).

The above models with metastable vacua may constitute a hidden sector in which supersymmery is broken. Then, the effects of supersymmetry breaking will be mediated to the
observable sector by gravity giving rise to soft supersymmetry breaking masses.
In the case of F--term supersymmetry breaking (in a metastable vacuum) squark and slepton masses may arise from nonrenormalizable Kahler potential terms like
$$\int d^4 \theta~ \bar Q_i Q_i \left(1+c_4 {{\Phi \bar \Phi} \over M_P^2}+c_5 {{q_i \bar q_i} \over M_P^2}+ \ldots \right)  \eqno(27)$$
where $Q_i$ denotes the MSSM matter superfields and $c_4,c_5$ are constants of $O(1)$. The sfermion masses squared are then given by 
$$ m_s^2 \sim c_4 {|F_{\Phi}|^2 \over M_P^2}+ c_5 {|F_{q_i}|^2 \over M_P^2} \eqno(28)$$ 
Gaugino masses, on the other hand, may arise from terms like
$$\int d^2 \theta~ Tr(W_aW_a) \left({1 \over g^2}+c_6 {\Phi \over M_P}+\ldots \right) \eqno(29)$$
where $W_a$ denotes the MSSM gauge superfields. Thus, the gaugino masses are given by $m_{\lambda} \sim c_6 F_{\Phi}/M_P$. 

Combining retrofitting in eq. (26) with the sfermion and gaugino mass terms above we can obtain semi--realistic phenomenology. We will not investigate the phenomenology of each of the 
metsstable nonsupersymmetric vacua we found
above. Rather, we will use only a couple of them to show that they can lead to semi--realistic phenomenology for reasonable choices of parameters.

Consider first the model of F--term supersymmetry breaking with $q_1=q_2$ in the metastable vacua given by eq. (13). In order to take the limit $m>>M>>\sqrt{F}$ as above, we
assume that each nonrenormalizable term in eq. (26) arises from a different gauge group with a different gaugino condensation scale $\Lambda_{1,2,3}$ with $\Lambda_1>\Lambda_3>>\Lambda_2$.
Using eqs. (3)-(5) we get the nonzero F--terms 
$$F_{q} \sim (m^3M^5)^{1/4} \qquad F_{\Phi} \sim -mM \eqno(30)$$
where we neglected all coefficients such as $c_i$ and $\lambda$. We see that $F_q<<F_{\Phi}$. The sfermion and gaugino masses are given by
$$m_s \sim m_{\lambda} \sim {{mM} \over M_P} \sim {{\Lambda_1^3 \Lambda_3^3} \over M_P^5} \eqno(31)$$
The requirement for $m_s \sim m_{\lambda} \sim~ TeV$ can be satisfied by taking $\Lambda_1 \Lambda_3 \sim 10^{31}~GeV$, e.g. $\Lambda_1 \sim 10^{16}~GeV$ and $\Lambda_3 \sim 10^{15}~GeV$. 
The metastability of the vacuum requires $e^{S_I} >e^{150}$ (where the age of the universe is taken to be about $e^{150}$ secs) which can be easily guaranteed if 
$S_I \sim (1/\lambda^2)>150$. This requires a rather small Yukawa coupling $\lambda < 1/25$.

A different metastable vacuum, e.g. the one with $q_1=q_2$ in the limit $\sqrt{F}>>M>>m$ which is given by eq. (14) can be obtained from only one gauge group with scale $\Lambda$. 
In this case we find
$$m_s \sim m_{\lambda} \sim {F \over M_P} \sim {\Lambda^3 \over M_P^2} \eqno(32)$$
TeV scale scalar and gaugino masses are obtained for $\Lambda \sim 10^{13}~GeV$. Again, stability of the vacuum requires, $S_I \sim \lambda^{-4}>150$ which can be satisfied for
$\lambda < 0.1$.

In both of the above scenarios, it is hard to obtain an acceptable $\mu$--term. A renormalizable coupling such as $\Phi H_1 H_2$ gives rise to a very large $\mu$. A nonrenormalizable term like
$\Phi^2 H_1 H_2/M_P$ gives rise to a $\mu$ of the correct order of magnitude but also to a very large $B$--term. Another possibility is to prevent the couplings of $\Phi$ and $\Phi^2$
to $H_1H_2$ by a discrete symmetry. Then we can obtain $\mu$ by retrofitting so that 
$\mu \sim \Lambda^3/M_P^2$. This, requires yet another non--Abelian gauge group with $\Lambda \sim 10^{13}~GeV$ (except in the second scenario in which this is the scale for $\Lambda$).

We stress that in the above estimates for the scalar and gaugino masses we neglected all pure numbers such as the coefficients $c_i$ and $\lambda$ which can change these estimates
by more than an order of magnitude. In fact, we saw that metastablity of the vacua requires a small $\lambda$.
Clearly, these numerical factors can be absorbed into small changes in the scales $\Lambda_i$. It would be interesting to investigate the phenomenology
of these models in greater detail.

\bigskip
\centerline{\bf 4. Conclusions and Discussion}
\medskip

In this paper, we constructed simple models with metastable nonsupersymmetric vacua in which supersymmetry is broken by nonzero F or D--terms at tree level. These results seem quite 
robust since they hold for a wide range of parameters. The models may have other metastable vacua in addition to the ones we found since our analysis was not complete 
due to our assumption that either $q_1=q_2$ or $q_1=-q_2$. 
We also showed that these models can lead to semi--realistic phenomenology by estimating the soft supersymmetry breaking masses in the metastable vacua. This is possible after retrofitting 
the models which leads to exponentially small masses and F--terms. Perhaps,
most importantly, these models can easily be embedded into larger ones with non--Abelian gauge groups and charged and neutral matter. They would especially play an important role at 
high energies where the physics is weakly interacting. 

Our analysis was done in the context of global supersymmetry. It would be interesting to find out whether our results remain valid in supergravity with the scalar potential
$$V=e^{K/M_P^2} \left(|\partial_i W+ \partial_i K {W \over M_P^2}|^2-3{|W|^2 \over M_P^2}\right)+g^2 |D|^2 \eqno(33)$$
The question is whether the nonsupersymmetric metastable vacua, which necessarily shift, nevertheless remain stable in supergravity. Even though supergravity corrections are Planck 
suppressed, due to the $-3 W^2/M_P^2$ term in eq. (33), the metastable vacua may be destabilized if the VEVs are large enough. 
In fact, in general, it is not even clear whether the globally supersymmetric vacua remain (with small shifts) stable in supergravity. This is guaranteed when $W=0$ in a 
supersymmetric vacuum but this is not the case for the our vacua (7) or (8). It would be interesting to repeat our analysis 
in supergravity and check if our results remain valid. 

As already mentioned in the introduction, the models above can be obtained by deforming ${\cal N}=2$ supersymmetric models (with only the Yukawa coupling in the superpotential). 
These deformations consist of the mass terms for the charged and singlet fields, F--terms for the singlets and anomalous D--terms. Deformed ${\cal N}=2$ supersymmetric models are 
easy to construct in string theory, for example in intersecting brane models[\GK] or in models obtained from branes wrapped on singularities[\SING]. Our results imply 
that such brane models will also have metastable nonsupersymmetric vacua in which supersymmetry is broken at tree level. (For some string or brane
constructions that lead to metastable vacua see [\BGH-\VER].) However, retrofitting which is crucial for semi--realistic phenomenology cannot be easily realized in intersecting brane 
constructions.
The only possibility for obtaining exponentially small parameters seems to be through geometric transitions[\GEO] in models with branes wrapped on singularities[\SING]. 

Any model with a supersymmetric vacuum that is parametrically far from a nonsupersymmetric one can in principle be used for inflation. A very flat scalar potential which is 
necessary for slow--roll
inflation can be obtained by making sure that the vacua are very far from each other. In general, there is a small parameter in the model (in our models this is the Yukawa coupling 
$\lambda$) which can be tuned to a small enough value to provide slow--roll inflation. For F--term supersymmetry breaking, this inflationary scenario is possible in supersymmetry 
but not in supergravity due to the well--known inflaton mass or $\eta$ problem. The alternative is D--term inflation[\DINF] which requires a model very similar to the one in section 
2.2. In fact, for small (or vanishing)
masses the model in section 2.2 is the same as the original D--term inflation models. Similarly, brane constructions of the above models may be good candidates for models of inflation if
they can be compactified. This rules out models based on intersecting branes but not those with branes at singularities[\EDI]. 


\bigskip
\centerline{\bf Acknowledgements}

I would like to thank the Stanford Institute for Theoretical Physics for hospitality.

\bigskip
\bigskip 
\centerline{\bf Appendix}

In this appendix, we elaborate on how to obtain the nonsupersymmetric metastable vacua in section 2.1 given by eqs. (14)-(21). 
In addition, we describe how to find the the metastability conditions for these nonsupersymmetric vacua.

The model with the superpotential in eq. (1) has a scalar potential given by ($V=V_F+V_D$ where $V_F$ and $V_D$ are given by eqs.
(2) and (6)) 
$$V=|\lambda q_1 q_2+2M \Phi +F|^2+ |(\lambda \Phi+m)q_2|^2+ |(\lambda \Phi+m)q_1|^2+ g^2(|q_1|^2-|q_2|^2)^2 \eqno(A.1)$$
In order to find the nonsupersymmetric metastable vacua of this potential we need to 
find all the critical points at which, $F_i \not =0$, i.e. supersymmetry is broken. Thus, we need to solve the minimization
conditions
$${\partial V \over \partial \Phi}=2M(\lambda {\bar q_1}{\bar q_2}+2M {\bar \Phi}+F)+\lambda ((\lambda {\bar \Phi}+m){\bar q_2})
+\lambda ((\lambda {\bar \Phi}+m){\bar q_1})=0 \eqno(A.2)$$
$${\partial V \over \partial q_1}=\lambda q_2(\lambda {\bar q_1}{\bar q_2}+2M {\bar \Phi}+F)-{\bar q_1}|\lambda {\bar \Phi}+m|^2=0
\eqno(A.3)$$
$${\partial V \over \partial q_2}=\lambda q_1(\lambda {\bar q_1}{\bar q_2}+2M {\bar \Phi}+F)-{\bar q_2}|\lambda {\bar \Phi}+m|^2=0
\eqno(A.4)$$
The general solution to these equations is quite complicated. However, we can assume that the VEVs are real in which case 
D--flatness requires $q_1= \pm q_2$.
If we assume $q_1=q_2$ we obtain the vacua in eqs. (9) and (10). On the other hand, if we take $q_1=-q_2$ we get the vacua in eqs. 
(16) and (17).

The scalar mass squared matrix is given by 
$$m_0^2=\pmatrix{
\overline {W}_{ac}W_{cb}& \overline{W}_{abc}W_c \cr
W_{abc} {\overline {W}_c}& W_{ac} \overline {W}_{bc} \cr
} \eqno(A.5)$$
where $W_a=\partial W/\partial q_a$, $\overline {W}_a=\partial \overline {W}/\partial \overline {q_a}$ etc.
Using the superpotential given by eq. (1) in $m_0^2$ we find the elements (in the basis ($q_1,q_2,
\Phi, {\bar q_1}, {\bar q_2}, {\bar \Phi}$))
$$(m_0^2)_{11}=\overline {(m_0^2)}_{44}=|\lambda \Phi+m|^2+|\lambda q_2|^2  \eqno(A.6)$$ 
$$(m_0^2)_{22}=\overline {(m_0^2)}_{55}=|\lambda \Phi+m|^2+|\lambda q_1|^2  \eqno(A.7)$$
$$(m_0^2)_{33}=\overline {(m_0^2)}_{66}=|\lambda q_2|^2+ |\lambda q_2|^2 +4M^2  \eqno(A.8)$$
$$(m_0^2)_{14}=(m_0^2)_{41}=(m_0^2)_{25}=(m_0^2)_{52}=(m_0^2)_{36}=(m_0^2)_{63}=0  \eqno(A.9)$$
$$(m_0^2)_{12}=\overline {(m_0^2)}_{21}=\overline {(m_0^2)}_{45}=(m_0^2)_{54}=\lambda^2 {\bar q_1} q_2  \eqno(A.10)$$
$$(m_0^2)_{13}=\overline {(m_0^2)}_{31}=\overline {(m_0^2)}_{46}=(m_0^2)_{64}=\lambda(\lambda {\bar \Phi}+m)q_1+ 
2 \lambda M {\bar q_2} \Phi  \eqno(A.11)$$
$$(m_0^2)_{23}=\overline {(m_0^2)}_{32}=\overline {(m_0^2)}_{56}=(m_0^2)_{65}=\lambda(\lambda {\bar \Phi}+m)q_2+ 
2 \lambda M {\bar q_1} \Phi  \eqno(A.12)$$
$$(m_0^2)_{15}=(m_0^2)_{24}=\overline {(m_0^2)}_{42}=\overline {(m_0^2)}_{51}=\lambda F+2 \lambda M \Phi \eqno(A.13)$$
$$(m_0^2)_{16}=(m_0^2)_{34}=\overline {(m_0^2)}_{43}=\overline {(m_0^2)}_{61}=\lambda^2 q_1 \Phi+\lambda m q_1  \eqno(A.14)$$
$$(m_0^2)_{26}=(m_0^2)_{35}=\overline {(m_0^2)}_{53}=\overline {(m_0^2)}_{62}=\lambda^2 q_2 \Phi+\lambda m q_2  \eqno(A.15)$$

A vacuum is metastable if $det(m_0^2)>0$ there. This guarantees that all scalars have positive masses squared
and therefore all directions in field space are locally stable. In order to check the metastability of a vacuum,
e.g. that given by eqs. (11) and (12) (or others given by eqs. (13) or (14)), we need to substitute the relevant VEVs in $(m_0^2)_{ij}$ given by the elements in eqs. (A.6)-(A.15) and look for the conditions under which $det(m_0^2)>0$.

\vfill

\refout

\end
\bye